# NUMERICAL SIMULATIONS OF THE EFFECT OF LOCALISED IONOSPHERIC PERTURBATIONS ON SUBIONOSPHERIC VLF PROPAGATION

*DESANKA ŠULIĆ[1], ALEKSANDRA NINA[1] AND VLADIMIR A. SREĆKOVIĆ[1]*

[1]*Institute of Physics, Belgrade Serbia*
e-mail: dusulic@ipb.ac.rs

**Abstract.** Electron density and temperature changes in the D-region of the ionosphere are sensitively manifested as changes in the amplitude and phase of subionospheric Very Low Frequency (VLF) signals propagating beneath the perturbed region. Disturbances (either in electron density or temperature) in the D region cause significant scattering of VLF waves propagating in the earth-ionosphere waveguide, leading to measurable changes in the amplitude and phase of the VLF waves. We analyze Lightning-induced electron precipitation (LEP) events during period 2008 – 2009 at Belgrade station on subionospheric VLF signals from four transmitters (DHO/23.4 kHz, Germany; GQD/22.1 kHz, UK; NAA/24.0 kHz USA and ICV/20.9 kHz Italy).

## 1. INTRODUCTION

One of the properties of the electromagnetic waves is that they reflect from conducting boundaries and can be guided between these boundaries. The surface of the Earth is moderately good conductor of the electricity and can reflect radio signals in the lower frequency signal range, including VLF. The ionosphere, on the other hand, is a complicated highly lossy (conducting) the anisotropic medium and D region is a good reflector of VLF waves. The Earth and the ionosphere constitute two boundaries of a waveguide within which VLF radio waves can propagate commonly referred to as the *earth-ionosphere waveguide*.

The propagation of electromagnetic signals in the earth-ionosphere waveguide can be thought in the terms of signals reflecting back and forth the earth and ionosphere. An electromagnetic wave totally reflects from a medium with varying dielectric properties at the point at which the reflective index is zero. For isotropic lossless plasma this condition is realized when $\omega \cong \omega_p$, where $\omega_p$ is the (angular) plasma frequency of the medium given by $\omega_p = 4\pi N_e e^2 / (\varepsilon_0 m_e)$, with $N_e$ being the number density of electrons, $e$ being the charge of an electron, $\varepsilon_0$ being the permittivity of free space and $m_e$ being the mass of an electron. However, in the VLF range the absorption, reflection and transmition of a radio signal incident on the ionosphere depends on factors such as: *the wave frequency, the angle of incidence, the altitude profiles of electron and ion concentrations, the altitude profile of electron temperature* and thus the rate at which these constituents collide with neutrals (i.e. the collision frequency), and the intensity and local orientation of the Earth's magnetic field. Due to the influence of the Earth's magnetic field the presence of absorption due to the collisions, the refractive index does not reach zero at any D-region altitude, so the total reflection never occurs, but substantial partial reflection is possible. The region of the ionosphere where such partial reflection occurs can be thought as a region where the reflection index changes very rapidly over distances comparable to a wavelength. When this occurs the region acts as a sharp boundary between two media and reflection occurs. A convenient quantity to describe the characteristics of the lower ionosphere is the "conductivity parameter" $\omega_r$ which is defined by:

$$\omega_r = \frac{\omega_p^2}{\nu} \cong 2{,}5 \cdot 10^5 \ \text{s}^{-1} \tag{1.1}$$

where $\omega_p$ is plasma frequency and $\nu$ is effective collision frequency of electrons and heavy particles (Wait and Spies, 1964). The reflecting properties of the ionosphere are thus dependent on the number density of electrons, $N_e$, which varies with height above the Earth surface and which is particularly variable in the nighttime D-region. A typical profile of the electron density in the critical altitude range from 60 to 90 km, can be approximately described by two-parameter exponential profile:

$$N(h) = \nu(h) \cdot 78{,}57 \cdot e^{\beta(h-h')} \quad (1.2)$$

with $N_e(h)$ is in electrons/m$^3$, *h'* being the effective reflection height in km and $\beta$ in km$^{-1}$ determining the sharpness or slope of the profile. The typical values of *h'* and $\beta$ for nighttime condition are *h'* = 85 km and $\beta = 0{,}50$ km$^{-1}$ respectively. Absorption of the radio waves in the ionosphere occurs primarily due to collision of electrons with neutral constituents of the atmosphere. The theoretical and the experimental profiles of collision frequency are nearly exponential and for nighttime condition the analytical form is:

$$\nu(h) = 1{,}86 \cdot 10^{11} e^{-0{,}15 \cdot h} \, [\text{s}^{-1}].$$

LEP events are produced by the fraction of the VLF energy radiated by lightning discharges that escapes into the magnetosphere and propagates as a whistler-mode wave. The whistler–mode wave interacts with trapped radiation belt electrons via cyclotron resonance, resulting in a pitch angle scattering of the electrons. If the pitch angle of radiation belt electrons at the edge of the loss cone is sufficiently decreased via resonant interaction with the whistler wave field, its lowered mirror height lies in the dense upper atmosphere and the particle is lost (i.e., precipitated) from the radiation belts.

## 2. VLF Remote Sensing

In this paper we restrict our attention to consider ionospheric disturbances which occur during nighttime when the ambient ionization levels are substantially lower and the ionospheric density enhancements associated with LEP events result in larger fractional changes in the overall ionospheric density, and hence the ionospheric disturbances are easier to detect via VLF remote sensing. VLF remote sensing is uniquely suited for the investigation of the nighttime D-region because of the sensitivity of subionosphericlly propagating VLF signals to changes in lower ionospheric conductivity.

VLF radio waves (~3 to 30 kHz) are guided by the spherical earth-ionosphere waveguide and can efficiently propagate to long distances. The amplitude and phase of the subionospherically propagating VLF signals depend sensitively on the electric conductivity change the amplitude and/or phase of VLF transmitter signals propagating in the earth-ionosphere waveguide on Great Circle Paths (GCPs) that pass through or near the localized disturbance. [Poulsen et al., 1993b]

Precipitating energetic electrons (induced by lightning-generated whistlers or other sources) deposit energy into the atmosphere and through secondary ionization alters the electron density and conductivity of the lower ionosphere. This ionospheric density enhancement in turn perturbs subionospheric VLF signals propagating on CGPs that pass through or near the disturbance. The amplitude and phase of VLF transmitter signals observed at any point can thus used to measure the spatial and temporal characteristics of localised disturbances in the lower ionosphere, a technique referred to as subionspheric VLF remote sensing.

## 3. Experimental Setup

Subionospheric signals from VLF transmitters were recorded on AWESOME ELF/VLF receiver system at Belgrade station. Data are typically acquired everyday at 18 to 06 UT when the GCPs between the transmitter and receiver are partially or whole in the nighttime sector. Two loops magnetic antennas are connecting to the preamplifier to detect VLF signals at all receivers. The phase and amplitude of the signals were logged with time resolution of 0,02 s. On Figure 1. is given case of 12 May 2009, when during whole night were recorded LEP events on two VLF signals: DHO/23.4 kHz $(53.10^0\,\text{N}, 7.60^0\,\text{E})$ Germany and GQD/22.1 kHz $(52.915^0\,\text{N}\ 3.28^0\,\text{W})$ UK recorded at Belgrade $(40.85^0\,\text{N}, 20.38^0\,\text{E})$. All LEP events presented on Figure 1 are manifested by increasing amplitude and decreasing phase.

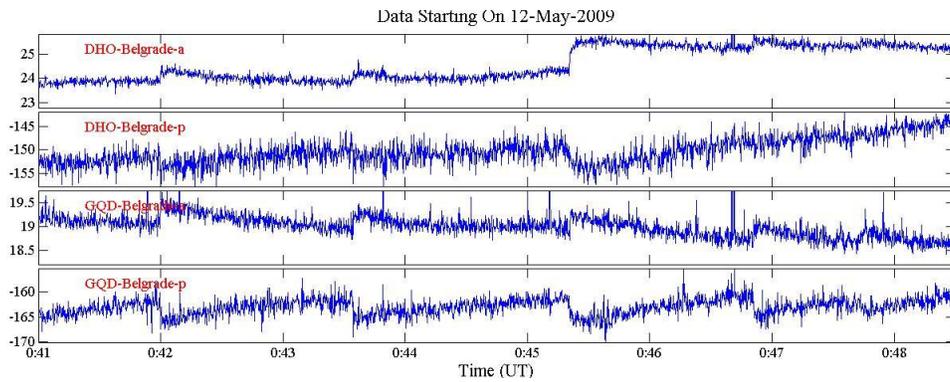

**Figure 1**. A 7:30 min snapshot of the received VLF signals: DHO/23.4 kHz and GQD/22.1 kHz, with typical LEP events

**Case: 12 September 2008** - LEP events on DHO/23.4 kHz transmitter between 01:00 and 03:00 on 12 September 2008 were recorded. Over the two-hour period 10 LEP events with perturbations of at least 1dB were recorded. Top panel on Figure 2 illustrates 32 min snapshot of recorded data. A large LEP event at ~02:36:50 UT is illustrated at lower panel. This LEP event is used for further analysis.

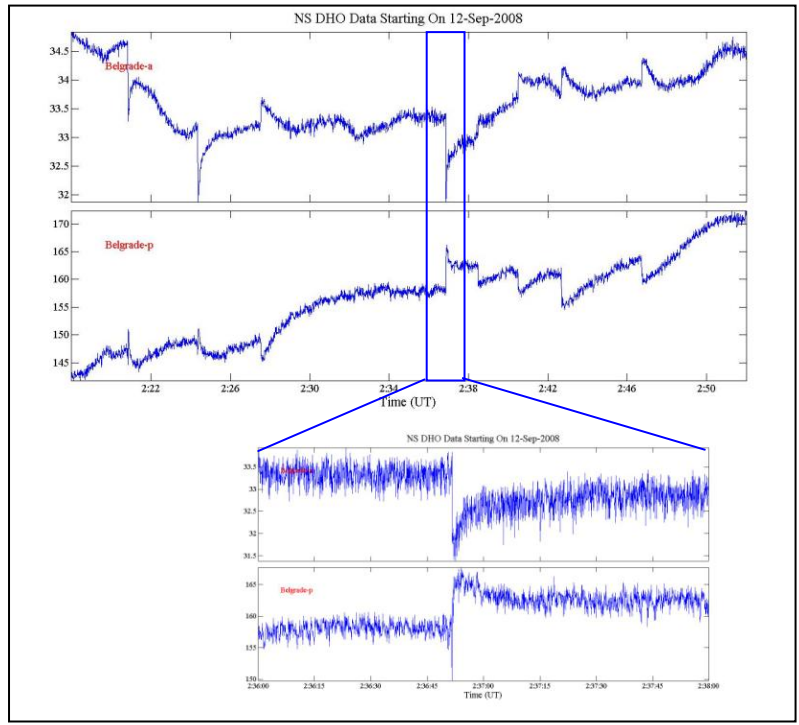

**Figure 2**. a) A 32 min snapshot of the received VLF signal (DHO/23.4 kHz) with typical LEP events b) an example of LEP event at 02:36:50 UT

In typical LEP events, the measurable features of $\Delta A_{rec}$ of the VLF signal refers to the change in amplitude measured in dB, from the ambient levels prior to the event, to the maximum (or minimum) levels reached during the event. The associated phase change $\Delta \phi_{rec}$ is also measured. The observed VLF amplitude and phase perturbations are simulated by the computer program Long-Wavelength Propagation Capability (LWPC), [Ferguson, 1998] using Wait's model of the lower ionosphere, as determined by two parameters: the sharpness $\beta$ and reflection height $h'$. By varying the values of $\beta$ and $h'$ so as to match the observed amplitude change $\Delta A_{rec}$ and phase change $\Delta \phi_{rec}$, the variation of the D-region electron density height profile $N_e(h)$ was reconstructed, throughout LEP events.

## 4. RESULTS

Method of varying the values of $\beta$ and $h'$ so as to match observed amplitude and phase perturbations is used for the approximate location and size of the associated ionospheric perturbation over transmitter–receiver GCP. Also with properly combination of $\beta$ and $h'$ as input parameters the electron density profile in the altitude range 60–90 km can be calculated. Generally it was found that during LEP events the reflection height $h'$ is lower than 87 km, which is the reflection height for regular nighttime condition. In our calculations it is between 87 and 85.3 km. The sharpness $\beta$ during LEP events is found in range 0.50 – 0.48 km$^{-1}$

The transmitter–receiver distance along the GCP for DHO/23.4 kHz Rhauderfehn (53.10 N, 7.60 E) to Belgrade (44.85 N, 20.38 E) is $D$-1304 km

|  | Recorded data | Modeling data |
|---|---|---|
| Amp. change | $\Delta A_{rec} = -1,4 dB$ | $\Delta A_{num} = -1,5 dB$ |
| Phase change | $\Delta \phi_{rec} = +9^0$ | $\Delta \phi_{num} = +8^0$ |

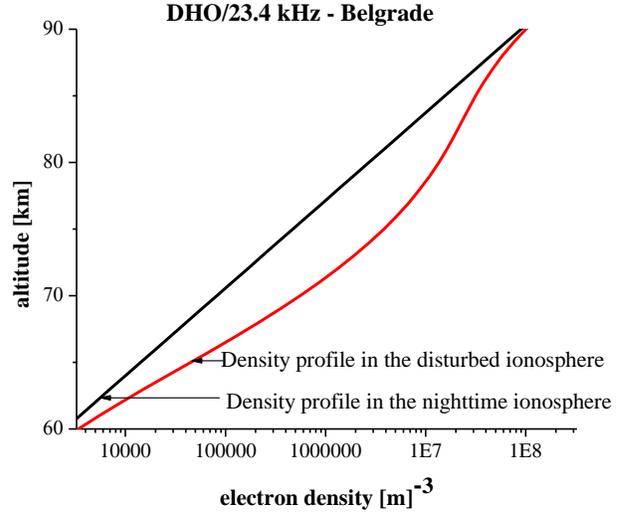

Figure 3. Profile of electron density

The corresponding change in electron density from the ambient value at the unperturbed reflection height, i.e. $N_e(87\text{km})=3.14\cdot 10^7 \text{el/m}^3$ to the value induced by LEP event up to $N_e(85.3\text{km})=3,67\cdot 10^7\ \text{el}/\text{m}^{-3}$ is obtained. The changes in $N_e(h)$ caused by LEP events within altitude range $h = 60$ to 90 km is determined and presented on the Figure 3. The recorded signals from transmitters in Europe are good base for studying localized ionization enhancements in the nighttime D region. By comparing simulated effects of LEP produced ionospheric disturbances on VLF signals with experimental data we are able to access the ionospheric electron density profiles most likely to have been in effect during the observed events.

## Referents
1. Ferguson J.A., *Computer Programs for Assessment of Long-Wavelength Radio Communications,* Version 2.0 Technical Document 3030, May 1998, Space and Naval Warfare Systems Center, CA 92152-5001, USA
2. Poulsen, W.L., Bell, T., Inan U.S., *The scattering of VLF waves by localized ionospheric disturbances produced by lightning induced electron precipitation,* Jour. of Geophysical Research, 15553-15559, 1998
3. Wait J.R. and Spies K.P. *Characteristics of the Earth-ionosphere waveguide for VLFF radio waves*, National Bureau of Standards, Technical Note, No 300, Issued December 30, 1964, Boulder, USA